\documentclass[12pt]{article}
\usepackage{graphicx}
\usepackage{epstopdf}
\usepackage{amssymb,amsmath,epsfig}
\usepackage{cite}
\usepackage[english]{babel}
\usepackage{array}
\usepackage[autostyle]{csquotes}
\begin{document}
\title{\bf Study of Kiselev Black Hole in Quantum Fluctuation Modified Gravity via Quasinormal modes and Greybody Factors}
\author {W. Sajjad \thanks{sajjadwaseem835@gmail.com }
and M. Azam \thanks{azam.math@ue.edu.pk, azammath@gmail.com},
\\
Department of Mathematics,\\ Division of Science and Technology,\\ University of Education, Lahore, Pakistan.}
\date{}
\maketitle
\begin{abstract}
In this work, we formulate the wave function to compute the effective potential of scalar and vector 
fields for the Kiselve black hole in the background of quantum fluctuation modified gravity.
Using the 6th-order Wentzel-Kramers-Brillouin approach, the quasinormal modes for the considered black hole 
are obtained through scalar and vector perturbation fields. We investigate the quasinormal frequency 
by considering several values of the metric and equation of state ($\mho$ and $\bar{\bar{\omega}}$)
parameter, respectively. The real part values of the quasinormal modes drop as the $\mho$ values rise. 
Later, the greybody factors of the black hole are determined and found that the parameters $\mho$, $\bar{\bar{\omega}}$ 
and the multipole moment $l$ significantly affects the greybody factors. Additionally, for scalar perturbations, 
we graphically analyze the rigorous bound of the greybody factor under the effects of parameters $\mho$ and $l$.
\end{abstract}

\section{Introduction}

Albert Einstein's general theory of relativity (GR) has made great strides in explaining observations 
and predicting spectacular events. However, it is still unable to link cosmic events or explain gravitational 
interaction. The universe is expanding at an accelerated rate, according to recent observational data. 
Gauss-Bonnet gravity, $f(R)$ theories, and $f(R, T)$ theories are more effective in explaining this behavior. 
In gravitational theories, black hole (BH) solutions are a significant study topic because they provide a 
vital platform for evaluating our comprehension of general relativity and modified gravity. Black holes 
are not solitary cosmic bodies; rather, they are dynamic objects that actively interact with their 
surroundings and significantly affect the structure of spacetime. Disturbing a BH causes ripples in 
spacetime itself, which gradually go away when the system stabilizes. Comprehending the way complex 
systems react to minor disturbances is essential to physics and offers a profound comprehension of spacetime and gravity.

Recently, Heisenberg's nonperturbative quantization \cite{1,2,3} was used to suggest that quantum fluctuation modified 
gravity (QFMG): If the metric can be decomposed into the total of the classical and quantum fluctuations 
portions, then the equivalent of quantum Einstein gravity at the classical level generates modified gravity 
models with a non-minimal coupling between geometry and matter \cite{4}.
This concept has been applied in several unique models, as seen in \cite{5,6,7,8}. Quantum fluctuation 
modified gravity exhibits a link between geometry and matter. It is anticipated that when the matter-geometry 
coupling is present, there will be variations between the solutions in QFMG and GR. In this approach, it is 
reasonable to examine the impacts imposed by QFMG on the scale of compact objects, as such variations may become 
more noticeable for high densities. The topic of BH solutions in QFMG has been discussed in \cite{9}.
Additionally, Friedmann-Lemaître-Robertson-Walker cosmology has been investigated in the framework of QFMG in 
references \cite{4,5,6,7,10}. Recently, baryogenesis within the context of QFMG was studied in \cite{11}. 
In \cite{11*}, an interesting BH solution of QFMG has been obtained, which
has two characteristic parameters besides mass: one parameter characterizes the quantum
fluctuation of metric, the other characterizes the matter surrounded the BH.

According to recent studies on black holes (BHs) conducted by the Event Horizon Telescope (EHT) team, the massive BH 
at the center of our galaxy possesses a complex structure of matter surrounding its event horizon \cite{12}. Investigating how 
quasinormal modes behave in different BH spacetimes such as those influenced by varying gravitational conditions or 
additional fields can enhance our understanding of the fundamental characteristics of BHs and their role in gravitational wave astronomy.
Traditionally, the oscillation modes of BHs have attracted significant interest from both gravitational theorists and 
experimentalists in the field of gravitational waves. These modes represent resonant, non-radial perturbations 
of BHs, similar to those of the Earth or the Sun, and can be excited by external disturbances. For instance, a 
binary BH merger can generate gravitational waves. The resulting signal can be categorized into three distinct 
phases, i.e., inspiral, merger, and ringdown. During the ringdown phase, the black hole emits gravitational waves 
known as quasinormal modes (QNMs), which possess unique and intricate frequencies \cite{13}.

When a BH stabilizes after being disrupted, its oscillatory behavior is described by complex frequencies 
called QNMs. These modes are distinct signals that rely on the BHs intrinsic 
characteristics, on BH parameters instead of how they are activated, and on the type of perturbation, whether 
it be gravitational, electromagnetic, or scalar. Quasinormal modes are particularly helpful 
for studying the characteristics of BHs because they offer a direct link between theoretical predictions 
and are independent of initial conditions. Quasinormal frequencies, also referred to as the ``characteristic sound" 
of BHs, have an imaginary part that indicates the damping rate and a real part that indicates the oscillation 
frequency. Note that the BH parameters, the classic ``fingerprints" of BHs, 
and the type of perturbation field are the only factors that affect QNMs. Improving our understanding of 
BHs can be achieved by solving the wave equation with specific boundary conditions, such 
as QNMs. Vishveshwara \cite{14} was the first to notice that the signal, which originates from a 
disrupted BH, decays with an exponential curve during the majority of the period.

Chandrasekhar and Detweiler conducted a comprehensive study on the QNMs of Schwarzschild black holes \cite{15}.
As a result, QNMs have gained more interest from researchers in recent years. Quasinormal modes of black holes 
are typically examined using the Wentzel-Kramers-Brillouin (WKB) approximation method \cite{16,17,18,19}. With 
the ongoing expansion of gravitational wave astronomy, understanding QNMs has become essential for interpreting 
data from detectors such as LIGO and Virgo \cite{20,21}.
Taking into account a range of perturbation parameter values, Fernando and Correa investigated the complete 
frequency spectrum of the QNMs of the regular BHs scalar field\cite{22}. In the eikonal limit, they have 
also calculated the scalar QNMs using the BHs unstable null geodesics. 
The investigation of greybody factors (GFs) provides a useful addition to the study of QNMs by clarifying 
the changes in the radiation spectrum when it leaves the gravitational pull of a BH. 
The greybody factor is essential for understanding the energy distribution of radiation emitted by BHs 
and affects the signals detected by gravitational wave observatories.

Furthermore, it was shown that the ringdown signal after an excessive mass ratio merger was significantly 
affected by the greybody component \cite{23}. The greybody factor describes the transmission probability 
of an outgoing wave reaching infinity or an incoming wave being absorbed by a BH\cite{24,25,26}.
A thorough grasp of BH dynamics and their interactions with the environment is possible 
when the greybody factor and QNMs are combined. Kokkotas et al. \cite{28} calculated Hawking radiation 
through gray-body components using the 6th-order WKB approach proposed by Konoplya \cite{27}. Furthermore, due 
to its versatility and adaptability, the 6th-order WKB technique is employed in many articles \cite{29,30,31} 
to produce Hawking radiation. In addition to the approximation, the GF can be found using the rigorous bound. A 
black hole can be described qualitatively using a rigorous bound. The QNMs, GFs, and rigorous bounds on 
GFs of BHs under QFMG are investigated in this paper. By examining these scenarios, we aim to gain insights 
into the behavior of black holes in complex electromagnetic environments and deepen our understanding of the universe's mysteries.

Our work is structured as follows: Section II offers a brief overview of the  BH line element in QFMG. In 
Section III, we examine the effective potential and the perturbations associated with the black hole. Sections IV and V 
analyze specific cases of rigorous bounds for the GFs and QNMs using the WKB method in the context of 
the black hole. Finally, a summary and conclusion are presented in Section VI.

\section{Black hole in modified gravity with quantum fluctuation}

In \cite{11*}, a significant quantum field theory modified gravity black hole (QFMGBH) solution to the gravitational 
field equations in QFMG, as proposed in \cite{4}, was derived. This solution can be simplified to characterize different 
types of black holes surrounded by Kiselev fluids by selecting specific values for the equation of state (EoS) parameter. 
In Schwarzschild coordinates, the geometry of the static and spherically symmetric QFMGBH is as follows:
\begin{equation}\label{1}
ds^{2}=-\textbf{Q}(r)dt^{2}+\frac{1}{\textbf{Q}(r)}dr^{2}+r^{2}(d\theta^{2}+sin^{2}\theta d\Phi^{2}),\\
\end{equation}
with
\begin{equation}\label{2}
\textbf{Q}(r)=1-\frac{2M}{r}+Pr^{-\frac{2(1+3\bar{\bar{\omega}}-4\bar{\bar{\omega}}\mho)}{2-3\mho+\bar{\bar{\omega}}\mho}},
\end{equation}
where \( P \) represents an integration constant, \( \mho \) denotes the parameter that describes the metric's 
fluctuations, and \( \bar{\bar{\omega}} \) indicates the equation of state (EoS) for Kiselev fluids. The black hole 
solution given by equation $(\ref{2})$ will reduce to the Schwarzschild black hole in the absence of fluid and 
fluctuations, where \( M \) is the total mass of the black hole. Our main objective is to study the impacts 
of the parameters \( \bar{\bar{\omega}} \) and \( \mho \) on the quasinormal modes and greybody factors. For 
various values of \( \mho \) and \( \bar{\bar{\omega}} \), the solution described by equation $(\ref{2})$ will 
correspond to different types of BH solutions, such as radiation, dust, quintessence, cosmological constant, or phantom.
\begin{figure}
    \centering
      {\includegraphics[width=0.45\textwidth]{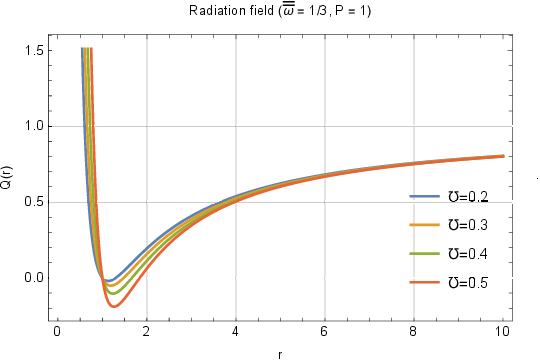}}
%   {\includegraphics[width=0.45\textwidth]{@Horizon1@.eps}}
   {\includegraphics[width=0.45\textwidth]{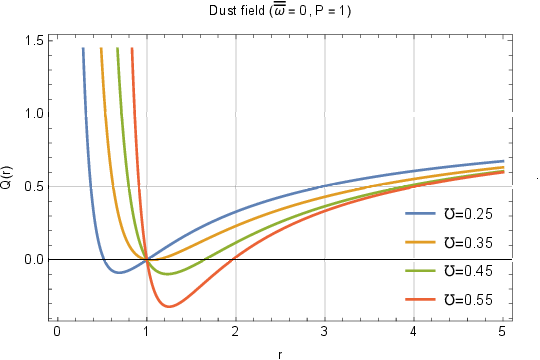}}
   {\includegraphics[width=0.45\textwidth]{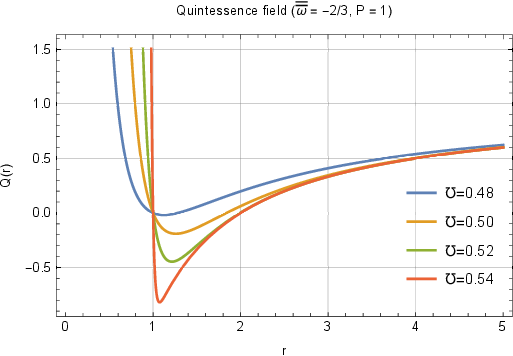}}
 {\includegraphics[width=0.45\textwidth]{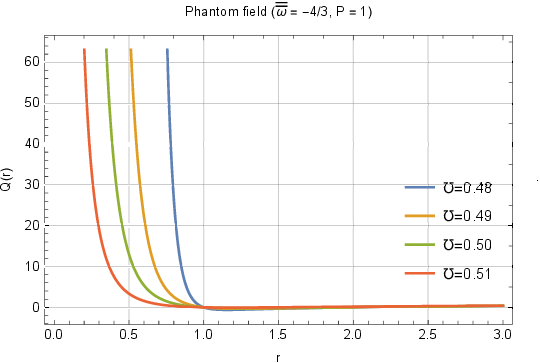}}
 {\includegraphics[width=0.45\textwidth]{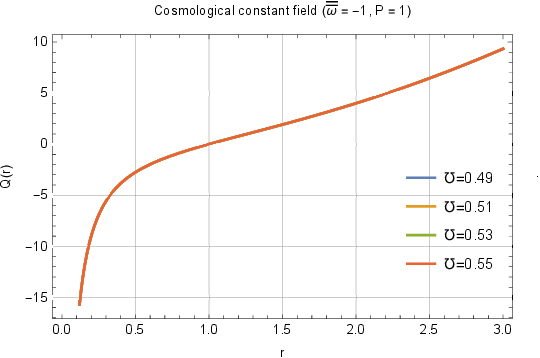}}
  \caption{The behavior of the black hole metric function concerning distance \( r \) for different metric fluctuation parameter values.}
    \label{fig:multi_graphs}\label{FIG.1}
  \end{figure} 
For a range of values of $\bar{\bar{\omega}}$ and \( \mho \), Figure $\textbf{1}$ shows the behavior of the metric 
function $(\ref{2})$. Different fields, such as the radiation field, dust field, and quintessence field, have this 
metric function defined in terms of the parameter \( r \). There are two event horizons for black holes in each of 
these domains. The event horizon's radius progressively grows as the values of the parameter \( \mho \) rise. When 
the parameter \( \mho \) in the phantom field is varied, the fourth plot demonstrates that the black hole maintains 
two event horizons. Also, this plot shows that when the metric parameter \( \mho \) grows, the horizons' radius 
shrinks. It is clear from the final figure that the black hole has a single event horizon throughout a range of 
values of \( \mho \) in the cosmological constant field; no other horizon is present. The final plot shows that 
the BHs inner horizon is unaffected by variations in the metric parameter values. 
  
\section{General formula for effective potential}

Master wave equations define the evolution of different fields within a given backdrop geometry and regulate the 
response of the BH spacetime to external disturbances. This section will analyze the massless scalar and vector 
perturbations of a static, spherically symmetric Kiskele BH in a QFMG.
Scalar and vector fields play a vital role in understanding BH physics, as they can represent matter fields, 
energy distributions, or perturbations. Moreover, their development offers important insights into the stability 
of BHs and the behavior of fields in varying geometries. The Klein-Gordon equation enhances our understanding of 
BH physics and the connection between GR and quantum mechanics. The Klein-Gorden (KG) equation is satisfied by 
a massless scalar field $\xi$ defined by
\begin{equation}\label{3}
\Box\xi\equiv \frac{1}{\sqrt{-\textrm{g}}}\partial_{\mu}\Big(\sqrt{-\textrm{g}}\textrm{g}^{\mu\nu}\partial_{\nu}\xi\Big)=0,
\end{equation}
where $\textrm{g}$ is the determinant of the metric tensor $\textrm{g}_{\mu\nu}$. 
The solution of the above equation can be obtain by the following choice of function
\begin{equation}\label{4}
\xi(t, r, \theta, \phi)=\frac{\textrm{\textit{e}}^{-i\hat{\omega}t}}{r}\breve{\xi}(r)\bar{Z}^{m}_{l}(\theta,\phi),
\end{equation}
here $\hat{\omega}$ is treated as frequency for the radial wave function ($\breve{\xi}(r)$). 
The azimuthal quantum number \( l \) is considered to be non-zero, while the quantum number \( m \) is set 
to zero. Consequently, Eq. (\ref{3}) results in a one-dimensional Schrödinger-like equation for the radial function \( \breve{\xi}(r) \):
\begin{equation}\label{15}
\frac{d^{2}}{dr^{2}_{*}}\breve{\xi}+(\hat{\omega}^{2}-\hat{V}_{eff})\breve{\xi}=0,
\end{equation}
with the ``tortoise" radial coordinate $r_{*}=\int\frac{dr}{\textbf{Q}(r)}$ and the effective potential $\hat{V}_{eff}$. 
For a scalar field with no mass, $\hat{V}_{eff}$ reads
\begin{equation}\label{16}
\hat{V}_{eff}(r)=\textbf{Q}(r) \Big[\frac{l(l+1)}{r^{2}}+\frac{1}{r}\frac{d\textbf{Q}(r)}{dr}\Big].
\end{equation}
For higher spin (boson) fields, the generalized form of effective potential has the form \cite{35,36}
\begin{equation}\label{17}
\hat{V}_{eff}(r)=\textbf{Q}(r) \Big[\frac{l(l+1)}{r^{2}}+(\tilde{\Lambda}-\tilde{\Lambda}^{2})\frac{1-
\textbf{Q}(r)}{r^{2}}+(1-\tilde{\Lambda})\frac{1}{r}\frac{d\textbf{Q}(r)}{dr}\Big],
\end{equation}
here, $\tilde{\Lambda}\leq l$ is the spin of the perturbative field and
\begin{center}
$
\tilde{\Lambda}=
\begin{cases}
0, & \text{scalar perturbation,}\\
1, & \text{electromagnetic perturbation. }  \\

\end{cases}
$
\end{center}
The effective potential $\hat{V}_{em}(r)$ for vector perturbations differs from that of 
the scalar perturbation $\hat{V}_{scalar}(r)$ because it incorporates the derivative of $\textrm{g}_{tt}$.

\subsection{Scalar and electromagnetic effective potential}

Here, we provide a brief overview of the properties of the perturb potentials related to the BH under 
consideration. Black hole QNMs are crucial because the nature of the perturb potential directly influences 
the frequencies and damping rates of these oscillatory modes. A preliminary understanding of the QNMs and 
their reliance on various parameters can be acquired by analyzing the behavior of the potential.\\
\begin{figure}[h!]
    \centering
   {\includegraphics[width=0.45\textwidth]{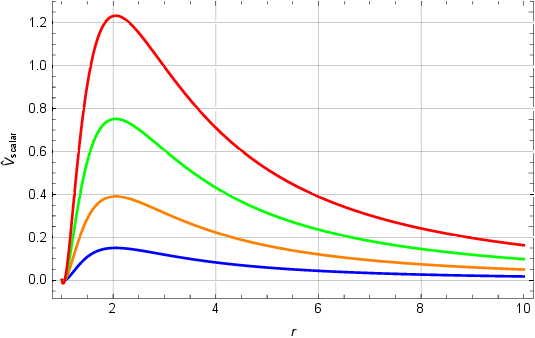}}\hfill
 {\includegraphics[width=0.45\textwidth]{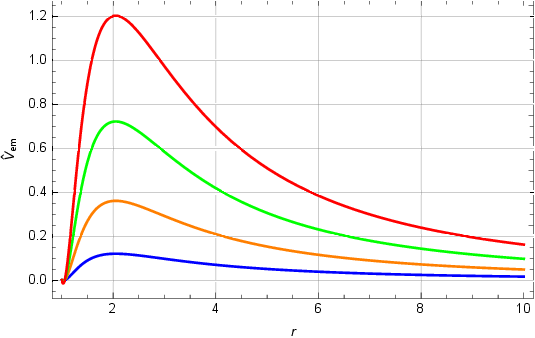}}\\
    \caption{The scalar potential $\hat{V}_{scalar}(r)$ and electromegnetic potential $\hat{V}_{em}(r)$ of a 
    Kiselve black hole vary with the radial distance $r$ in a quantum fluctuation-modified gravity 
    for different multipole moment values $l$,~ $l = 1~ blue~ line,~ l = 2 ~orange ~line, ~l = 3
     ~green~ line, ~l = 4~ red ~line$. where $M = 1$,~$P = 1$,~$\mho = 0.1$, and $\bar{\bar{\omega}} = 0.3$.}
    \label{fig:multi_graphs}\label{FIG.2}
\end{figure}
\
Figure $\textbf{2}$ displays the scalar potential (left panel) for a range of multipole moment $(l)$ values. 
As expected, the potential's peak value increases as $l$ increases. This pattern is consistent with the 
understanding that for higher values of multipole moments, which are linked to more complex angular 
dependencies in the perturbation, lead to a larger potential barrier. The rise in the potential peak 
with increasing $l$ indicates that the black hole is more robust to disturbances with higher angular 
momentum, which is likely to lead to QNMs with shorter durations and higher frequency. This happens 
because the potential barrier traps the disturbance closer to the black hole, accelerating 
the decay and increasing the oscillation frequencies.
\begin{figure}[h!]
    \centering
   {\includegraphics[width=0.45\textwidth]{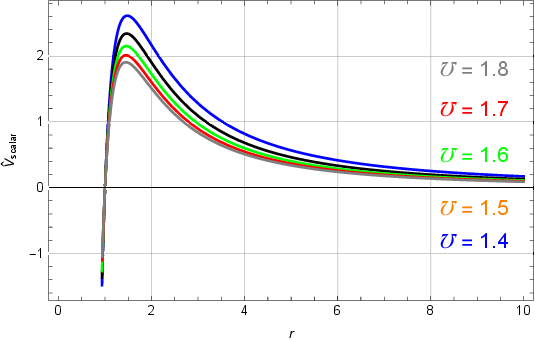}}\hfill
   {\includegraphics[width=0.45\textwidth]{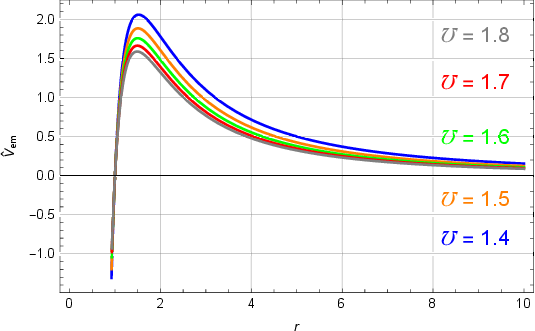}}\\
    \caption{For varying coupling constant $\mho$, the scalar potential $\hat{V}_{scalar}(r)$ and 
    electromegnetic potential $\hat{V}_{em}(r)$ of the Kiselve black hole in a 
    quantum fluctuation-modified gravity with the radial distance $r$, where $M = 1$,~$P = 1$,~~$l = 2$, and $\bar{\bar{\omega}} = 0.3$ are involved.}
    \label{fig:multi_graphs}\label{FIG.3}
\end{figure}
In Fig. $\textbf{3}$, the scalar effective potential is presented as a function of the radial coordinate \( r \) for various values 
of the parameter $\mho$. It is clear that as the value of $\mho$  increases from $1.4$ to $1.8$, the height of the barrier 
decreases gradually. This indicates that the scalar field encounters a lower barrier when $\mho$ is higher. 
From the perspective of QNMs, a lower barrier potential has two significant effects:\\
$1$. oscillation frequency decreases: Because the barrier is weaker, the trapped perturbations oscillate at a lower frequency.
$2$. Damping rate slows down: The perturbations take longer to dissipate, which means that the black hole ``rings” for a longer 
period before fading out. Therefore, the figure illustrates that increasing $\mho$ reduces the BHs resistance to scalar 
perturbations, resulting in lower quasinormal mode (QNM) frequencies and slower decay rates.

Figure $\textbf{4}$ demonstrates that as the parameter $\bar{\bar{\omega}}$ increases, the scalar potential barrier 
decreases. This reduction means that the black hole becomes less resistant to scalar perturbations. As a result, the 
damping rate declines, and the QNM frequencies become lower, indicating that perturbations dissipate more slowly.
\begin{figure}[h!]
    \centering
   {\includegraphics[width=0.45\textwidth]{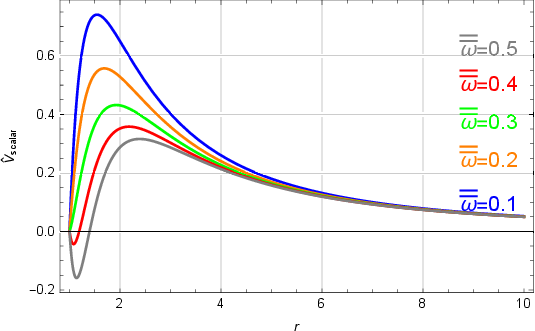}}\hfill
   {\includegraphics[width=0.45\textwidth]{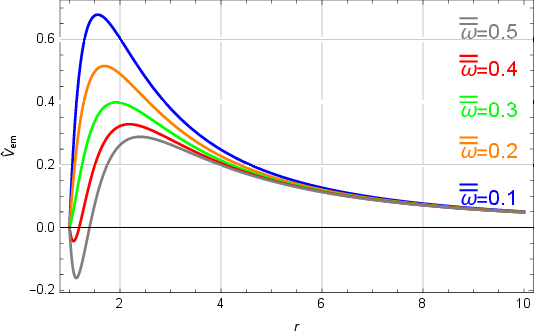}}\\
    \caption{In a quantum fluctuation-modified gravity with the radial distance $r$, the scalar potential 
    $\hat{V}_{scalar}(r)$ and electromegnetic potential $\hat{V}_{em}(r)$ of the Kiselve black hole vary 
    as the parameter of the equation of state of the fluid $\bar{\bar{\omega}}$ is varied, where $M = 1$,~$P = 1$,~$l = 2$,~$\mho = 0.01$.}
    \label{fig:multi_graphs}\label{FIG.4}
\end{figure}
Taking into consideration different parameters and multipole moment values, similar curves have also been 
created for the electromagnetic potential (right panel) in Figs. $\textbf{2}$, $\textbf{3}$, and $\textbf{4}$. The qualitative 
behavior of the scalar potential is similar to that of the electromagnetic potential curves. The main 
difference is that the graph values of scalar perturbation are greater than those of electromagnetic 
perturbation for the same constant values. This disparity suggests that the BH spacetime and the 
scalar field interact more strongly, which increases the potential barriers. For scalar perturbations, 
we expect the QNM spectrum to be higher frequency and shorter duration than for electromagnetic disturbances. 
The QNMs associated with electromagnetic perturbations may have lower frequencies and longer damping durations, 
indicating a more progressive decay of these modes, as indicated by the lower potential values in the electromagnetic scenario.
The analysis of the perturbation potential provides valuable insights into the nature of the QNMs for both scalar 
and electromagnetic perturbations. Changes in the potential concerning the multipole moment $l$, the parameters 
$\bar{\bar{\omega}}$, and  $\mho$ show how sensitive the QNMs are to different parameters. This information is 
crucial for predicting the stability and dynamical response of BHs to perturbations, particularly when 
working with different gravity models or the presence of other fields.

\section{Quasinormal modes with special cases using the WKB technique }

Here, we estimate the QNMs of BHs using the WKB methodology, a popular approximation method 
in the field of BH perturbation theory. Schutz and Will \cite{18} introduced the WKB technique, which provides 
the first-order approximation for computing QNMs. Despite its usefulness, the approach is known to have certain 
limitations, including a higher degree of error in specific scenarios. In response to these limitations, scholars 
have developed higher-order WKB approximations, which significantly improve the accuracy of QNM calculations. 
As a result, the QNM in the ringdown phase exhibits damped oscillations with discrete complex frequencies. 
The QNM frequency imaginary and real components correspond to the decay time and the perturbation's oscillation frequencies, 
respectively. Because QNMs only depend on the BH parameters and are independent of the initial perturbation, 
they are essential tools for studying BHs and gravity theories.
There are a number of different ways to obtain QNM frequencies. To study scattering around BHs, however, 
we will employ the 6th WKB approximation \cite{27}, which was first presented in \cite{19}, to increase the 
efficiency of computing the quasinormal frequencies. Gravitational waves primarily exist in the basic mode characterized 
by \( l = 2 \) and \( \tilde{y} = 0,1 \). Therefore, we will concentrate on this mode to examine how the 
parameter \( \mho \) affects the quasinormal frequencies. Here is the complex frequency formula in the 6th-order WKB approximation

\begin{equation}\label{18}
i\frac{(\hat{\omega}^{2}-\hat{V}_{0})}{\sqrt{-2\hat{V}^{\prime\prime}_{0}}}-\sum_{t=2}^{6}\digamma_{t}=\tilde{y}+\frac{1}{2},
\end{equation}
where the maximum effective potential is $\hat{V}_{0}$. The formula 
$\hat{V}^{\prime\prime}_{0}=\frac{d^{2}\hat{V}(r_{*})}{dr^{2}_{*}}\Big|_{r_{*}=r_{1}}$, ~~$r_{1}$ here is
the position where the effective potential is largest, and $\digamma_{t}$ represents the corrections term 
for the 2nd to 6th order WKB \cite{19,27}.

In the following subsections, we consider different special cases corresponding to 
different values of the parameter $\bar{\bar{\omega}}$ and investigate the effects of
$\mho$ in QFMG on the QNMs. Theoretically, any value can be assigned to the state's parameter $\bar{\bar{\omega}}$. 
However, we examine some specific values that are often chosen due to their significant physical consequences.
Tables $1$, $2$, $3$, $4$, and $5$ show that there is a negative imaginary part in each obtained BH frequency, which indicates stability. 
To observe the impact of the model parameters on the QNM spectrum, the real (R) and imaginary (IM) QNMs are plotted 
clearly in terms of the model parameters. The behavior of the graphs is quite similar for cases $1$ through $4$, 
as illustrated graphically. The real and imaginary components of the quasinormal modes are depicted as a 
function of the parameter $\mho$ for massless scalar and electromagnetic disturbances in Figs. $\textbf{5}$ to $\textbf{8}$. 
Increasing $\mho$ causes a large drop in the real part of the QNMs, which represents the oscillation frequency. 
At the same time, the imaginary component increases as $\mho$ increases, indicating a slowdown in the damping rate. 
In general, the system oscillates weakly and decays slowly.
The variation of real and imaginary QNMs concerning model parameter $\mho$ for massless scalar and electromagnetic 
perturbations is shown in Fig. $\textbf{9}$. Unlike traditional tendencies, the real component of the QNMs remains 
nearly constant, hence the oscillation frequency does not decrease as $\mho$ increases. Additionally, the 
imaginary part of the QNMs does not seem to be increasing, suggesting that the damping rate is almost constant. 
This is in contrast to previous studies, which found that as $\mho$ increased, frequency decreased and damping time changed relatively small.

\textbf{Case 1:}\\
For the case of radiation field, we take $\bar{\bar{\omega}} = 1/3$, cosequently, $\textbf{Q}(r)$ has the form
\begin{equation}\label{19}
\textbf{Q}(r)=1-\frac{2M}{r}+Pr^\frac{-12+8\mho}{6-8\mho}.
\end{equation}
The QNMs of the massless scalar and electromagnetic perturbation with a range of $\mho$ values are displayed 
in Table $1$ along with the overtone number ($\tilde{y} = 0, 1$) and assuming the other parameters are $M = 1$,~$l = 2$,~$P = 1.1$.\\
\begin{figure}[h!]
    \centering
   {\includegraphics[width=0.45\textwidth]{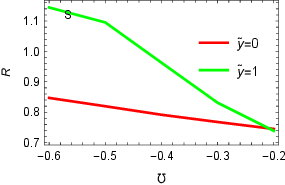}}\hfill
 {\includegraphics[width=0.45\textwidth]{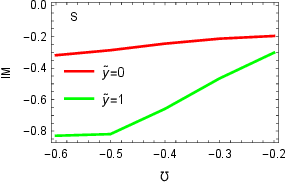}} \\
  {\includegraphics[width=0.45\textwidth]{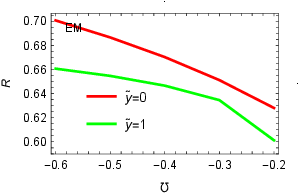}} \hfill
{\includegraphics[width=0.45\textwidth]{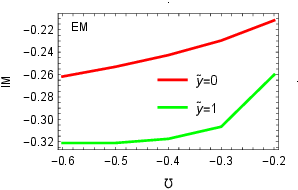}} 
    \caption{The QNM frequencies' real and imaginary components are shown in Table 1 as a function of 
    the parameter $\mho$. Above is a scalar disturbance, while below is an electromagnetic disturbance. 
    $M = 1$,~$l = 2$,~$P = 1.1$,~and $\bar{\bar{\omega}} = 1/3$ are present in this case.}
    \label{fig:multi_graphs}\label{FIG.5}
\end{figure}
\begin{table}[h!]
\caption{\label{tab1}The BH QNMs for specific $\mho$ values for both scalar and electromagnetic disturbances. 
In this case, $M = 1$,~$l = 2$,~$P = 1.1$,~and $\bar{\bar{\omega}} = 1/3$.}
\centering
    \small % Table ko chhota karne ke liye
    \renewcommand{\arraystretch}{0.7} % Row height ko adjust karne ke liye
    \setlength{\tabcolsep}{7pt} % Column width ko adjust karne ke liye
    \resizebox{1\textwidth}{35pt}{ % Puri table ki width kam karne ke liye
    \begin{tabular}{c|cc|cc}
    \hline
     & \multicolumn{2}{c|}{\textbf{Scalar perturbation}} & \multicolumn{2}{c}{\textbf{Electromagnetic perturbation}} \\
    \hline
    \textbf{$\mho$} & \textbf{$\tilde{y}$=0} & \textbf{$\tilde{y}$=1} & \textbf{$\tilde{y}$=0} & \textbf{$\tilde{y}$=1} \\
    \hline
-0.6 & 0.8470-0.3178i & 1.1446-0.8295i &0.7007-0.2618i &0.6607-0.3211i \\ 
-0.5 & 0.8196-0.2860i & 1.0954-0.8192i &0.6866-0.2531i & 0.6547-0.3211i  \\ 
-0.4 & 0.7916-0.2442i & 0.9629-0.6586i &0.6703-0.2427i & 0.6466-0.3174i \\ 
-0.3 &  0.7677-0.2127i & 0.8311-0.4657i &0.6512-0.2297i & 0.6346-0.3066i \\
-0.2 & 0.7449-0.1954i & 0.7387-0.3017i&0.6281-0.2117i & 0.6013-0.2602i\\
\hline
 \end{tabular}
    }
 \end{table}

\textbf{Case 2:}\\
For the case of dust field, where $\bar{\bar{\omega}} = 0$ and consequently, we have
\begin{equation}\label{20}
\textbf{Q}(r)=1-\frac{2M}{r}+Pr^{-\frac{2}{2-3\mho}}.
\end{equation}
Table $2$ represents the QNMs for electromagnetic and massless scalar perturbations with a range of $\mho$ values
and $M = 1$,~$l = 2$,~$P = 0.9$, and an overtone number ($\tilde{y} = 0, 1$) using the 6th-order WKB approximation.
\begin{table}[h!]
\caption{\label{tab2}For particular $\mho$ values, the QNMs of BH for both scalar 
     and electromagnetic perturbations. We have $M = 1$,~$l = 2$,~$P = 0.9$,~and $\bar{\bar{\omega}} = 0$ in this case.}
    \centering
    \small % Table ko chhota karne ke liye
    \renewcommand{\arraystretch}{0.7} % Row height ko adjust karne ke liye
    \setlength{\tabcolsep}{7pt} % Column width ko adjust karne ke liye
    \resizebox{1\textwidth}{35pt}{ % Puri table ki width kam karne ke liye
    \begin{tabular}{c|cc|cc}
    \hline
     & \multicolumn{2}{c|}{\textbf{Scalar perturbation}} & \multicolumn{2}{c}{\textbf{Electromagnetic perturbation}} \\
    \hline
    \textbf{$\mho$} & \textbf{$\tilde{y}=0$} & \textbf{$\tilde{y}=1$} & \textbf{$\tilde{y}=0$} & \textbf{$\tilde{y}=1$} \\
    \hline
    0.01  & 0.9049-0.3985i  & 1.1366-0.9485i  & 1.0865-0.1491i  & 2.1035-0.1306i  \\ 
    0.02  & 0.9006-0.3973i  & 1.1304-0.9425i  & 1.0576-0.1489i  &  2.0357-0.1301i  \\ 
    0.03 & 0.8956-0.3950i  & 1.1194-0.9312i  & 1.0272-0.1487i  & 1.9675-0.1296i  \\ 
    0.04 &  0.8902-0.3918i  & 1.1045-0.9152i  & 0.9949-0.1487i  & 1.8989-0.1290i  \\

    \hline
    \end{tabular}
    }
\end{table}
\begin{figure}[h!]
    \centering
   {\includegraphics[width=0.45\textwidth]{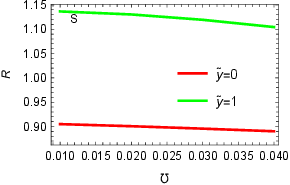}}\hfill
   {\includegraphics[width=0.45\textwidth]{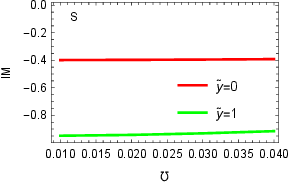}} \\
  {\includegraphics[width=0.45\textwidth]{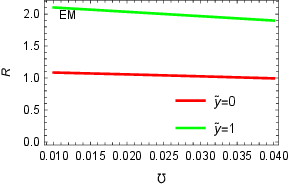}}\hfill
  {\includegraphics[width=0.45\textwidth]{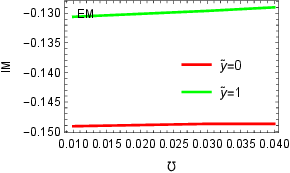}} 
    \caption{The QNM frequencies are presented with real and imaginary parts as a function of the parameter $\mho$. 
    Scalar disruption (above), electromagnetic disruption (below). Specifically, $M = 1$,~$l = 2$,~$P = 0.9$,~and $\bar{\bar{\omega}} = 0$.}
    \label{fig:multi_graphs}\label{FIG.6}
\end{figure}

\textbf{Case 3:}\\
For the case of quintessence field we take $\bar{\bar{\omega}} = - 2/3$, hence 
\begin{equation}\label{21}
\textbf{Q}(r)=1+\frac{-2M}{r}+Pr^\frac{-2(-3+8\mho)}{6-11\mho}.
\end{equation}
With this choice of $\textbf{Q}(r)$, Table $3$ similar to the above case represents the QNMs with the following assumption
of parametric values of the BH 
$M = 1$,~$l = 2$,~$P = 0.4$, and the overtone number ($\tilde{y}$) is $0, 1$. 
\begin{table}[h!]
\caption{\label{tab4}The QNMs of BH for specific $\mho$ values for both electromagnetic 
and scalar disturbances. We have $M = 1$,~$l = 2$,~$P = 0.4$,~and $\bar{\bar{\omega}} = - 2/3$ in this case.}
    \centering
    \small % Table ko chhota karne ke liye
    \renewcommand{\arraystretch}{0.7} % Row height ko adjust karne ke liye
    \setlength{\tabcolsep}{7pt} % Column width ko adjust karne ke liye
    \resizebox{1\textwidth}{35pt}{ % Puri table ki width kam karne ke liye
    \begin{tabular}{c|cc|cc}
    \hline
     & \multicolumn{2}{c|}{\textbf{Scalar perturbation}} & \multicolumn{2}{c}{\textbf{Electromagnetic perturbation}} \\
    \hline
    \textbf{$\mho$} & \textbf{$\tilde{y}=0$} & \textbf{$\tilde{y}=1$} & \textbf{$\tilde{y}=0$} & \textbf{$\tilde{y}=1$} \\
    \hline
    0.02 & 1.6397-1.3588i& 3.0160-2.9667i&0.9744-0.6103i & 3.6282-0.2445i     \\ 
  0.04 & 1.6200-1.3389i  & 2.9732-2.9237i &0.9744-0.6064i& 3.5797-0.2423i\\ 
    0.06 & 1.5993-1.3180i&2.9280-2.8783i&0.9729-0.6015i& 3.5299-0.2400i\\ 
   0.08& 1.5775-1.2958i&2.8805-2.8301i&0.9700-0.5955i& 3.4790-0.2375i\\
  0.1& 1.5544-1.2723i&2.8303-2.7789i& 0.9651-0.5881i & 3.4271-0.2348i\\
    \hline
    \end{tabular}
    }
    
\end{table}
\begin{figure}[h!]
    \centering
   {\includegraphics[width=0.45\textwidth]{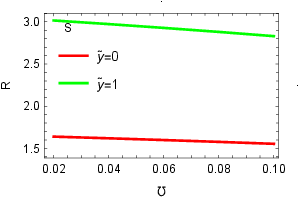}}\hfill
   {\includegraphics[width=0.45\textwidth]{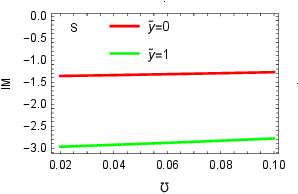}}\\
  {\includegraphics[width=0.45\textwidth]{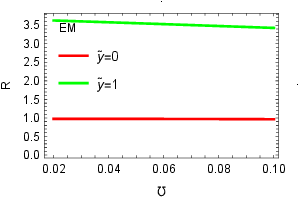}}\hfill
  {\includegraphics[width=0.45\textwidth]{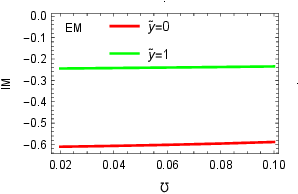}} 
    \caption{Real and imaginary parts of QNM frequencies are displayed as a function of the 
    parameter $\mho$ $(Table 4)$. Scalar disturbance (above) and electromagnetic disturbance (below). In this case, $M = 1$,~$l = 2$,~$P = 0.4$,~and $\bar{\bar{\omega}} = - 2/3$.}
    \label{fig:multi_graphs}\label{FIG.7}
\end{figure}

\textbf{Case 4:}\\
For the case of phantom field, we take $\bar{\bar{\omega}} = - 4/3$, as a consequence $\textbf{Q}(r)$ has the form
%%Case 5
%\subsection{QNMs of black hole in a quantum fluctuation-modified gravity encircled by a phantom field}
%Lastly, we investigate $\hat{\omega}=-4/3$. The following expression now provides the metric function $\textbf{Q}(r)$.
\begin{equation}\label{22}
\textbf{Q}(r)=1-\frac{2M}{r}+Pr^\frac{-2(-9+16\mho)}{6-13\mho}.
\end{equation}
The QNMs of the massless scalar and electromagnetic perturbation corresponding to with a range of $\mho$ values 
and $M = 1$,~$l = 2$,~$P = 0.01$ are shown in ~Table $4$.
\begin{figure}[h!]
    \centering
{\includegraphics[width=0.45\textwidth]{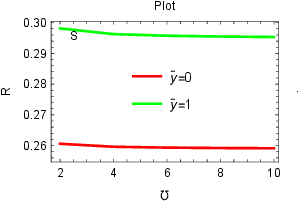}}\hfill
    {\includegraphics[width=0.45\textwidth]{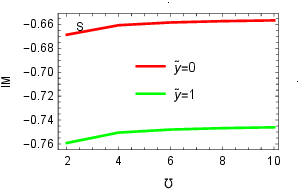}}\\
 {\includegraphics[width=0.45\textwidth]{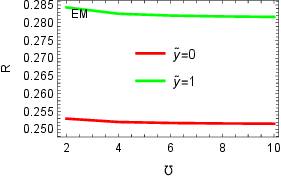}}\hfill
 {\includegraphics[width=0.45\textwidth]{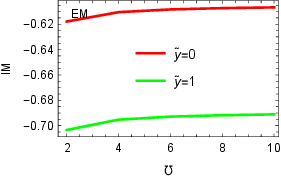}} 
    \caption{Real and imaginary parts of the QNM frequencies are displayed as a function of 
    the parameter $\mho$ (Table 5). Scalar disturbance (above), electromagnetic disturbance (below). 
    $M = 1$,~$l = 2$,~$P = 0.01$,~and $\bar{\bar{\omega}} = - 4/3$ are the values in this case.}
    \label{fig:multi_graphs}\label{FIG.8}
\end{figure}
\begin{table}[h!]
\caption{\label{tab4}The QNMs of BH for specific $\mho$ values for both electromagnetic and scalar disturbances. 
In this instance, $M = 1$,~$l = 2$,~$P = 0.01$,~and $\bar{\bar{\omega}} = - 4/3$.}
    \centering
    \small % Table ko chhota karne ke liye
    \renewcommand{\arraystretch}{0.7} % Row height ko adjust karne ke liye
    \setlength{\tabcolsep}{7pt} % Column width ko adjust karne ke liye
    \resizebox{1\textwidth}{35pt}{ % Puri table ki width kam karne ke liye
    \begin{tabular}{c|cc|cc}
    \hline
     & \multicolumn{2}{c|}{\textbf{Scalar perturbation}} & \multicolumn{2}{c}{\textbf{Electromagnetic perturbation}} \\
    \hline
    \textbf{$\mho$} & \textbf{$\tilde{y}=0$} & \textbf{$\tilde{y}=1$} & \textbf{$\tilde{y}=0$} & \textbf{$\tilde{y}=1$} \\
    \hline
    2& 0.2605-0.6687i& 0.2981-0.7591i&0.2529-0.6179i& 0.2843-0.7034i    \\ 
   4 & 0.2596-0.6608i  & 0.2962-0.7504i&0.2520-0.6105i& 0.2825-0.6952i\\ 
    6 & 0.2593-0.6585i& 0.2957-0.7479i&0.2517-0.6083i& 0.2820-0.6929i\\ 
 8 & 0.2592-0.6574i&0.2954-0.7467i&0.2516-0.6073i& 0.2817-0.6917i\\ 
  10 & 0.2591-0.6567i&0.2953-0.7460i&0.2515-0.6067i& 0.2816-0.6911i\\ 
    \hline
    \end{tabular}
    }
    \end{table}
    
\textbf{Case 5:}\\
Lastly, for the case of the cosmological constant field, we take $\bar{\bar{\omega}} = - 1$, 
accordingly $\textbf{Q}(r)$ has the form 
\begin{equation}\label{23}
\textbf{Q}(r)=1-\frac{2M}{r}+Pr^{2},
\end{equation}
and the metric solution in this case is the same as the one that Kiselev obtained in GR \cite{33}. In this case, 
~Table $5$ displays the QNMs of the massless scalar and electromagnetic perturbation  
by considering $M = 1$, ~$l = 2$, ~$P = - 0.01$, for different values of $\mho$.
\begin{figure}[h!]
    \centering
   {\includegraphics[width=0.45\textwidth]{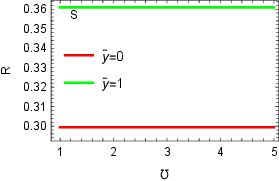}}\hfill
   {\includegraphics[width=0.45\textwidth]{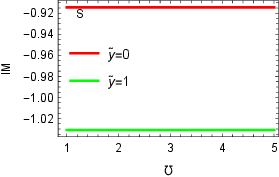}}\\
  {\includegraphics[width=0.45\textwidth]{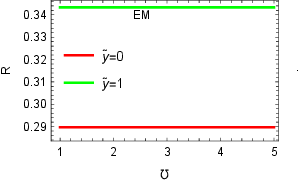}}\hfill
  {\includegraphics[width=0.45\textwidth]{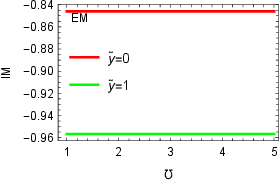}} 
    \caption{Plotting of QNM frequencies with real and imaginary parts as a function 
    of parameter $\mho$ is shown in Table 5. The scalar disturbance is above, and the 
    electromagnetic disturbance is below. Specifically, $M = 1$, ~$l = 2$, ~$P = - 0.01$, and ~$\bar{\bar{\omega}} = - 1$.}
    \label{fig:multi_graphs}\label{FIG.9}
\end{figure}
\begin{table}[h!]
\caption{\label{tab3}The BH QNMs for specific $\mho$ values for both scalar and electromagnetic disturbances. 
In this case, $M=1$,~$l=2$,~$P=-0.01$,~and~$\bar{\bar{\omega}} = - 1$.}
    \centering
    \small % Table ko chhota karne ke liye
    \renewcommand{\arraystretch}{0.7} % Row height ko adjust karne ke liye
    \setlength{\tabcolsep}{7pt} % Column width ko adjust karne ke liye
    \resizebox{1\textwidth}{35pt}{ % Puri table ki width kam karne ke liye
    \begin{tabular}{c|cc|cc}
    \hline
     & \multicolumn{2}{c|}{\textbf{Scalar perturbation}} & \multicolumn{2}{c}{\textbf{Electromagnetic perturbation}} \\
    \hline
    \textbf{$\mho$} & \textbf{$\tilde{y}=0$} & \textbf{$\tilde{y}=1$} & \textbf{$\tilde{y}=0$} & \textbf{$\tilde{y}=1$} \\
    \hline
    1  & 0.2994-0.9145i  & 0.3609-1.0309i& 0.2897-0.8460i & 0.3433-0.9567i         \\ 
    2  & 0.2994-0.9145i   & 0.3609-1.0309i & 0.2897-0.8460i & 0.3433-0.9567i \\ 
    3 & 0.2994-0.9145i & 0.3609-1.0309i& 0.2897-0.8460i  & 0.3433-0.9567i  \\ 
    4 & 0.2994-0.9145i  & 0.3609-1.0309i & 0.2897-0.8460i & 0.3433-0.9567i   \\
  
    \hline
    \end{tabular}
    }
    
\end{table}
The integration constant in the metric function \( \textbf{Q}(r) \) naturally arises during the solution 
of the field equations. These constants often have physical significance, representing key characteristics 
of BH spacetimes. In our analysis, we have made this constant adjustable for simplicity, enabling us to 
easily examine the impact of other parameters on the effective potential, GFs, and QNMs. 

\section{Greybody factors}

The concept of greybody factors originated with Hawking's groundbreaking results in $1975$, 
which demonstrated that black holes are not completely black but emit radiation, now known as 
Hawking radiation \cite{34}. Although this emission occurs near the BHs horizon, the spectrum 
of the radiation that is visible to a distant observer is altered by a redshift factor. 
This distortion, known as the greybody factor \cite{35,36}, reflects the difference between 
the original Hawking radiation and what an observer at infinity detects. Numerous methods, 
such as those made by Fernando \cite{38} and Maldacena et al. \cite{37}, can be used to calculate greybody factors.

\subsection{Studying of GFs Special cases by using the WKB approach}

We used the higher-order WKB approximation method to calculate GFs for scalar and 
electromagnetic disturbances. Specifically, this method helps to determine the transmission 
and reflection coefficients involved in wave scattering near a BH. Similar to scattering 
waves from a BHs horizon, we examine the wave equation under boundary conditions that 
allow incoming waves from infinity. This reveals how much of the incident wave is 
reflected and how much passes through the potential barrier. In this context, 
the transmission coefficient is defined as the GFs. We consider
\begin{equation}\label{24}
\underline{\psi}=\check{\textrm{\textit{e}}}^{-i\hat{\omega}r_{*}}+\textrm{\textit{R}}
\check{\textrm{\textit{e}}}^{i\hat{\omega}r_{*}}~~as~~r_{*}\rightarrow+\infty,~~~~
\underline{\psi}=\textrm{\textit{T}}\check{\textrm{\textit{e}}}^{-i\hat{\omega}r_{*}}~~as~~r_{*}\rightarrow-\infty,
\end{equation}
where $\textrm{\textit{T}}$ and $\textrm{\textit{R}}$ stand for the transmission and reflection coefficients, respectively. 
These coefficients satisfy the conservation relation $|\textrm{\textit{T}}|^{2}+|\textrm{\textit{R}}|^{2} = 1$, 
indicating that the sum of the transmission and reflection probabilities must be equal to one. 
The transmission coefficient $|\textrm{\textit{T}}|^{2}$, also known as the GFs $(\textbf{A})$, 
measures the proportion of radiation that crosses the potential barrier and reaches an observer at infinity; mathematically,
\begin{equation}\label{25}
|\textrm{\textit{T}}|^{2}=|\textbf{A}|^{2}=1-|\textrm{\textit{R}}|^{2}=|A|^{2}.
\end{equation}
Using the WKB approximation, the reflection coefficient $\textrm{\textit{R}}$ can be written as
\begin{equation}\label{26}
\textrm{\textit{R}}=\big(1+\check{\textrm{\textit{e}}}^{-2i\pi\widehat{\kappa}}\big)^{-\frac{1}{2}},
\end{equation}
where the phase factor $\widehat{\kappa}$ is represented as
\begin{equation}\label{27}
\widehat{\kappa}-i\frac{(\hat{\omega}^{2}-\hat{V}_{0})}{\sqrt{-2\hat{V}^{\prime\prime}_{0}}}-\sum_{t=2}^{6}\digamma_{t}=0.
\end{equation}
Although the WKB method is generally effective, its accuracy diminishes at low frequencies due to the 
tendency for total reflection, which causes the GFs to approach zero. However, this 
limitation does not impact the calculation of energy emission rates. 
A detailed review of this method is beyond the scope of our study, as the WKB 
approach is robust and widely applicable in various scenarios, including investigations of black hole 
disturbances and greybody effects. We refer readers to see \cite{39,40,41,42} for comprehensive study.

In this subsection, we examine how the multipole moment $l$ and metric parameter $\mho$ of QFMG
affect the GFs for a few exceptional situations that correspond to various values of the 
parameter $\bar{\bar{\omega}}$. In the equation of the state, the parameter $\bar{\bar{\omega}}$ 
can theoretically have any value. This article looks at many specific values that are selected due to their significant physical implications.
Similar to the cases of QNMs corresponding to different values of $\bar{\bar{\omega}}$. Here, we take the same values
i.e., $\bar{\bar{\omega}} = 1/3, 0, -2/3, -4/3,$ and $-1$ for the GFs.\\

Figures $\textbf{10}$, $\textbf{11}$, $\textbf{12}$, $\textbf{13}$, and $\textbf{14}$ provide a detailed analysis of the behaviour of the GFs
for massless scalar perturbation $|A_{s}|^{2}$ and electromagnetic perturbation $|A_{e}|^{2}$ under 
various model parameters. Each image shows how different factors influence the absorption probability, 
which offers valuable insights into BH dynamics in the context of QFMG. 
It is suggested that higher-energy modes are more strongly stimulated with larger $l$ values, 
as the peak of the greybody components shifts to higher frequencies $\hat{\omega}$ as the $l$ increases. 
It is interesting to observe that, in contrast to scalar perturbations, electromagnetic perturbations 
exhibit the maximum absorption levels at gradually decreasing frequencies. This suggests that 
electromagnetic waves interact better at lower frequencies than scalar waves, resulting in earlier 
absorption peaks for the same multipole moments. The significance of $\mho$ in controlling the 
interaction between BH and outside perturbations is demonstrated by the sensitivity of the GFs to this value. 
For both electromagnetic and scalar disturbances, the GFs exhibit a discernible decline with increasing $\mho$. 
Black holes with smaller $\mho$ values are thought to have higher absorption and scattering probability, 
which means they can efficiently absorb more entering particles or radiation. We also look at how the metric 
parameter $\mho$ affects greybody factors in Fig. $\textbf{14}$ (left panel). The current graph indicates that the 
greybody factor is unaffected by changes in $\mho$. This demonstrates that under the given conditions, 
BHs absorb and scatter incoming radiation in the same manner, independent of changes in $\mho$.
In the right panel of Fig. $\textbf{14}$, we can observe that for the scalar case with $l = 3$, the 
curve approaches a nearly straight line. In contrast, for the vector case, a similar trend appears 
at $l = 1$. This indicates that, at these specific values of multipole moments, the greybody factors 
exhibit little to no significant change. As a result, it becomes challenging to determine 
whether radiation is being emitted or absorbed at this point.

Concluding this section, it is demonstrated that the parameters $l$ and $\mho$ significantly influence the 
greybody factors for both scalar and electromagnetic disturbances. Moreover, electromagnetic disturbances 
typically absorb more at lower frequencies than scalar perturbations. These results improve our knowledge 
of the complex interactions between external disturbances and BHs in the context of QFMG.

\begin{figure}
 
   {\includegraphics[width=0.38\textwidth]{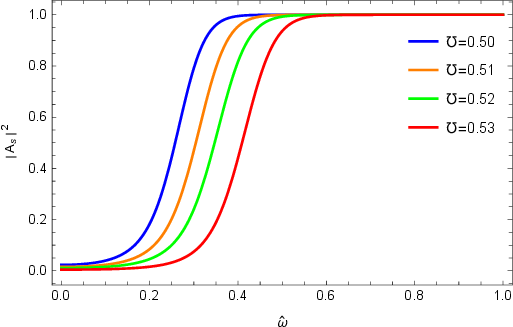}}\hfill
   {\includegraphics[width=0.38\textwidth]{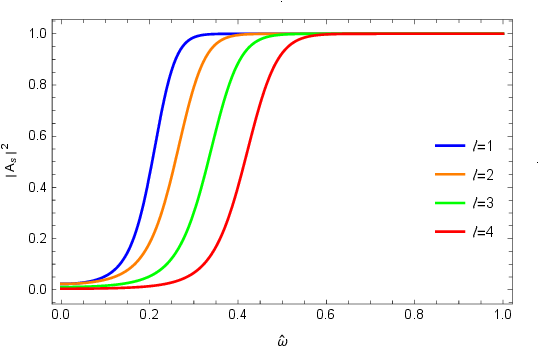}}\\
     {\includegraphics[width=0.38\textwidth]{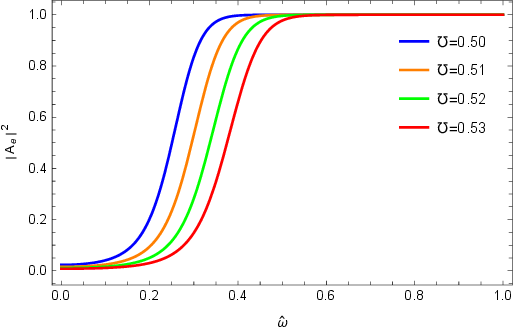}}\hfill 
   {\includegraphics[width=0.38\textwidth]{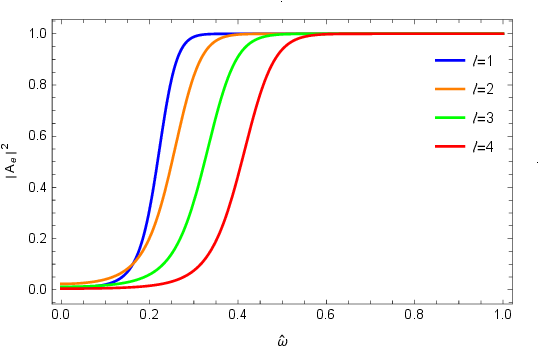}}\\
    \caption{Greybody factors for electromagnetic (Below) and massless scalar (Above) perturbations for various multipole moment $l$ and metric parameter $\mho$ values with parameter values, $M = 1$, $\bar{\bar{\omega}} = 1/3 $~(radiation field), and~$P = 1$.}
    \label{fig:multi_graphs}\label{FIG.10}
    
      {\includegraphics[width=0.38\textwidth]{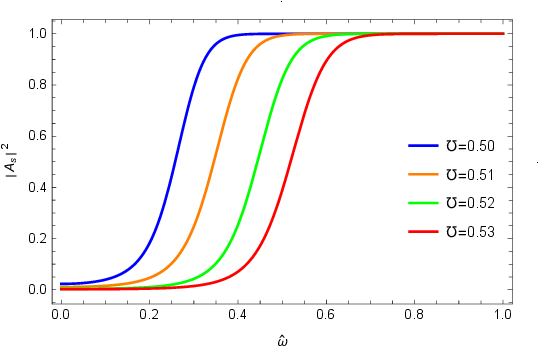}}\hfill
   {\includegraphics[width=0.38\textwidth]{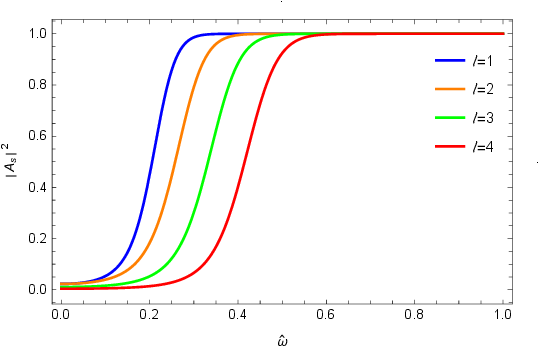}}\\
     {\includegraphics[width=0.38\textwidth]{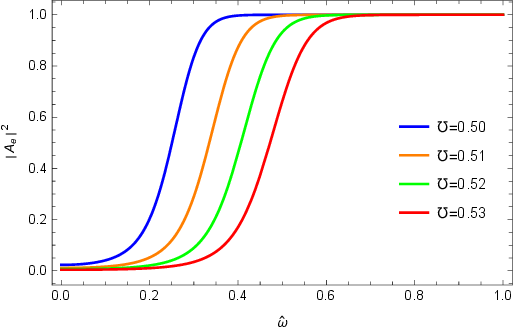}}\hfill 
   {\includegraphics[width=0.38\textwidth]{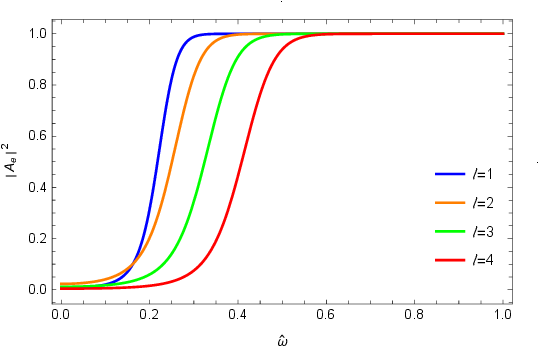}}\\
    \caption{Greybody factors for massless scalar (Above) and electromagnetic (Below) perturbations with parameter values $M = 1$, $P = 1$, and~ $\bar{\bar{\omega}} = 0$ ~(dust field) for different values of the multipole moment $l$ and metric parameter $\mho$.}
    \label{fig:multi_graphs}\label{FIG.11} 
 
   \end{figure}
\newpage
   \begin{figure}
       {\includegraphics[width=0.38\textwidth]{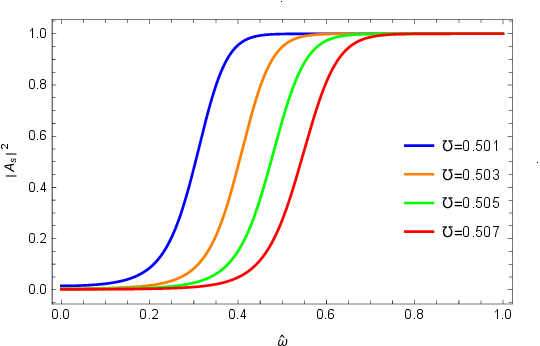}}\hfill
   {\includegraphics[width=0.38\textwidth]{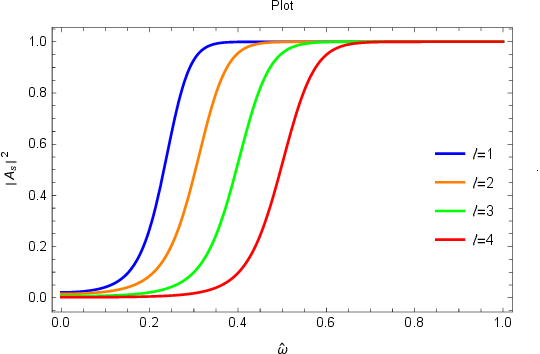}}\\
    {\includegraphics[width=0.38\textwidth]{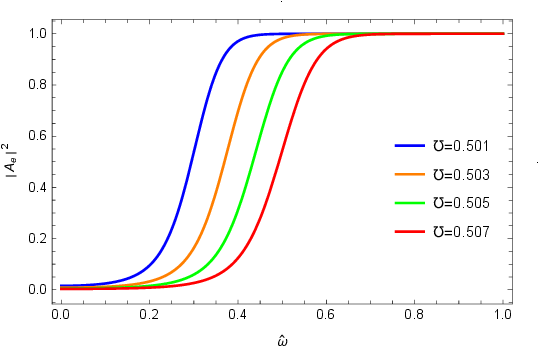}}\hfill
   {\includegraphics[width=0.38\textwidth]{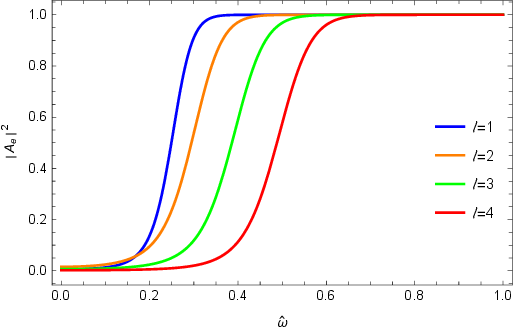}}\\
    \caption{Greybody factors with parameter values $M = 1$, $P = 1$, and~ $\bar{\bar{\omega}} = - 2/3 $~(quintessence field) for electromagnetic (Below) and massless scalar (Above) perturbations for different values of the multipole moment $l$ and metric parameter $\mho$.}
    \label{fig:multi_graphs}\label{FIG.12}
    
       {\includegraphics[width=0.38\textwidth]{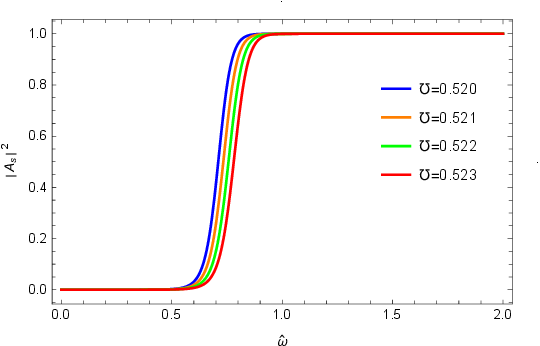}}\hfill
 {\includegraphics[width=0.38\textwidth]{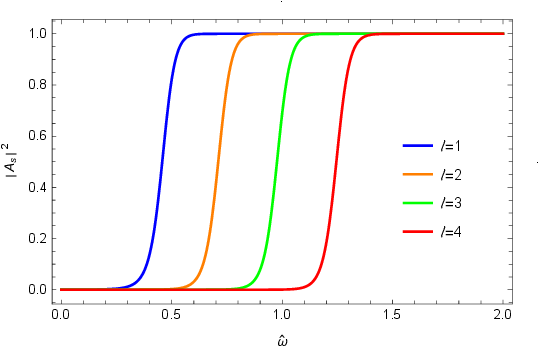}}\\
  {\includegraphics[width=0.38\textwidth]{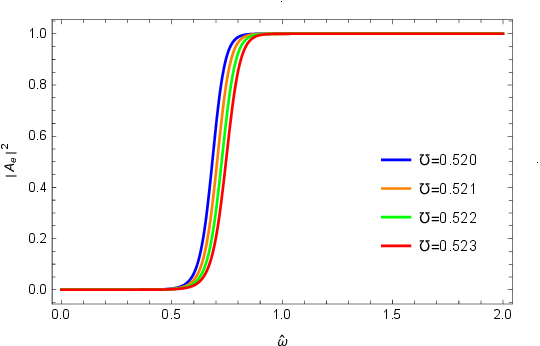}}\hfill
 {\includegraphics[width=0.38\textwidth]{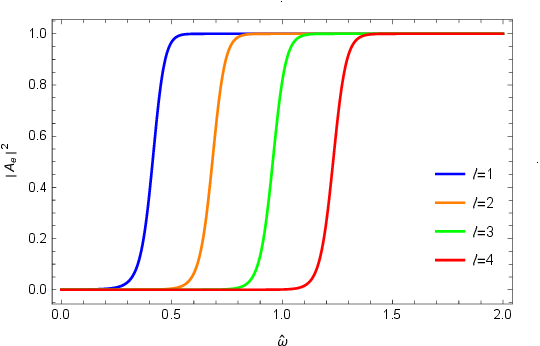}}\\
    \caption{Greybody factors for electromagnetic (Below) and massless scalar (Above) 
    disturbances with parameter values $M = 1$, $\bar{\bar{\omega}} = - 4/3$~(phantom field), and $P = 1$ with 
    varying multipole moment $l$ and metric parameter $\mho$ values.}
    \label{fig:multi_graphs}\label{FIG.13}
      \end{figure}
 \newpage 
      \begin{figure}
       {\includegraphics[width=0.38\textwidth]{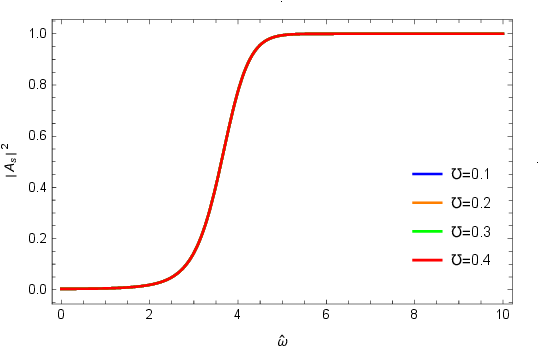}}\hfill
 {\includegraphics[width=0.38\textwidth]{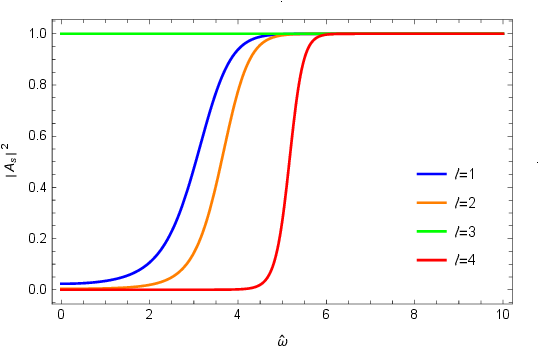}}\\
  {\includegraphics[width=0.38\textwidth]{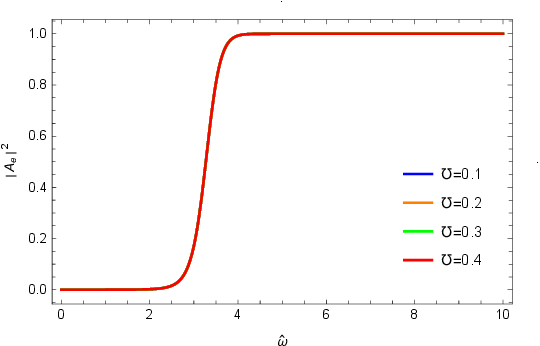}}\hfill
 {\includegraphics[width=0.38\textwidth]{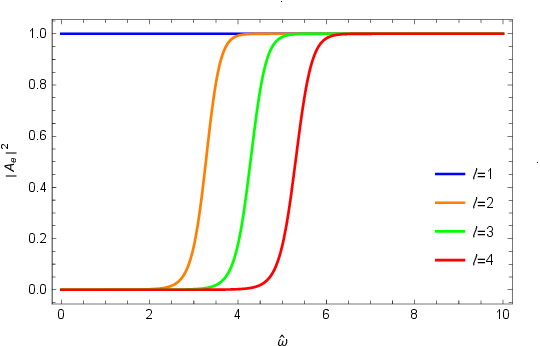}}\\
    \caption{Greybody factors for massless scalar (Above) and electromagnetic (Below) perturbations 
    with parameter values $M = 1$, $\bar{\bar{\omega}} = - 1 $~(cosmological constant field), and~$P = 1$ for different values of the multipole moment $l$ and metric parameter $\mho$.}
    \label{fig:multi_graphs}\label{FIG.14}   
\end{figure}

\subsection{Analysis of Special Cases of Rigorous Bounds on Greybody Factors}

In this section, we adopt a different strategy to examine rigorous bounds on greybody 
components, and it is limited to the study of scalar perturbations, since scalar and 
electromagnetic perturbations behave similarly regarding greybody components. Following 
the technique introduced by Visser \cite{45} and later developed by Boonserm and 
Visser \cite{46}, Boonserm et al. \cite{47}, Ngampitipan et al. \cite{48}, and 
others \cite{49,50,51,52,53}, the limits of the GFs of the BH in a QFMG have 
been formulated by using the following formula
\begin{equation}\label{28}
T_{a}\geq sech^{2}\bigg(\frac{1}{2\hat{\omega}}\int_{-\infty}^{+\infty}|\hat{V}_{eff}(r)| \frac{dr}{\textbf{Q}(r)} \bigg),
\end{equation}
here, the greybody component is represented by the transmission coefficient $T_{a}$, whereas the perturbation frequency is represented by $\hat{\omega}$,
and
\begin{equation}\label{29}
\hat{V}_{eff}(r)=\frac{l(l+1)\textbf{Q}(r)}{r^{2}}+\frac{\textbf{Q}(r)}{r}\frac{d\textbf{Q}(r)}{dr}.
\end{equation}
Boonserm et al. \cite{54} modify the boundary conditions to account for the 
coupling constant and the parameter of the equation of state.
The modified bound can be expressed as follows
\begin{equation}\label{30}
\hat{A}\geq T_{a}= sech^{2}\bigg(\frac{1}{2\hat{\omega}}\int_{r_{H}}^{R_{H}}|\hat{V}(r) |
 \frac{dr}{\textbf{F}(r)} \bigg)=sech^{2} \bigg(\frac{C_{l}}{2\hat{\omega}}\bigg),
\end{equation}
and the integral term $C_{l}$ has the following form
\begin{equation}\label{31}
C_{l}= \int_{r_{H}}^{R_{H}}\frac{\big|\hat{V}(r) \big| }{\textbf{Q}(r)}dr = \int_{r_{H}}^{R_{H}}\bigg|\frac{l(l+1)}{r^{2}}+\frac{\textbf{Q}^\prime}{r}\bigg| dr.
\end{equation}
Here, $r_{H}$ and $R_{H}$ corresponds to the BHs event and cosmic horizon radii, respectively.
Consequently, the bound of transmission coefficient for the considered object turns out to be
\begin{eqnarray}\label{32}\notag
T_{a}=sech^{2}\bigg[\frac{1}{2\hat{\omega}}\bigg(\frac{M+l(l+1)r_{H}-
\frac{2P(-1+(-3+4\mho)\bar{\bar{\omega}})r_{H}^{3+\frac{6(-1+\mho)(1+\bar{\bar{\omega}})}{2+\mho(-3+\bar{\bar{\omega}})}}}
{-4-6\bar{\bar{\omega}}+\mho(3+7\bar{\bar{\omega}})}}{r_{H}^{2}}-
\notag\\ \frac{M+l(l+1)R_{H}-
\frac{2P(-1+(-3+4\mho)\bar{\bar{\omega}})R_{H}^{3+\frac{6(-1+\mho)(1+\bar{\bar{\omega}})}{2+\mho(-3+\bar{\bar{\omega}})}}}
{-4-6\bar{\bar{\omega}}+\mho(3+7\bar{\bar{\omega}})}}{R_{H}^{2}}\bigg)\bigg],
\end{eqnarray}
where $\mho$, $P$, and $\bar{\bar{\omega}}$ represent the metric constant, integration constant, and the fluid's equation of state, respectively.
This expression provides a strict lower bound on the GFs as a function of various parameters. Our results 
demonstrate that the BH parameters that govern the gravity model determine the GFs bound. 
These findings will contribute to a better understanding of BH radiation and the interaction between BHs and quantum fluctuation-modified gravity.

We examined the greybody bounds in this subsection by focusing solely on scalar perturbations because the behavior 
of greybody components is similar for scalar and electromagnetic perturbations. The behavior of the rigorous bounds on 
greybody factors for massless scalar perturbations under different model parameters is detailed in Figs. $\textbf{15}$, $\textbf{16}$, $\textbf{17}$, $\textbf{18}$, and $\textbf{19}$. 
Similar to the cases of QNMs with different values of $\bar{\bar{\omega}} = 1/3, 0, -2/3, -4/3, and -1$, 
we used the same values for the rigorous bounds on GFs.
Numerical calculations allow us to evaluate the rigorous bound for the case of $l = 2$, which is shown 
in all figures for the case of different $l$ values (on the right panel) and different $\mho$ values (on the left panel).
\begin{figure}
 {\includegraphics[width=0.40\textwidth]{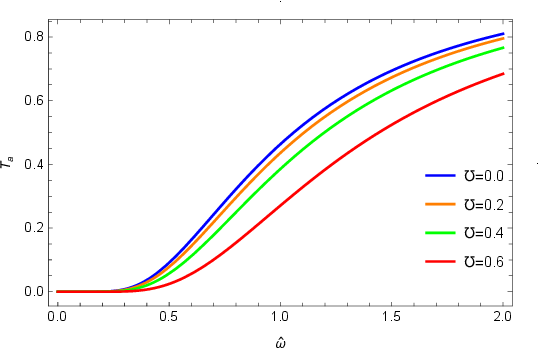}}\hfill
 {\includegraphics[width=0.40\textwidth]{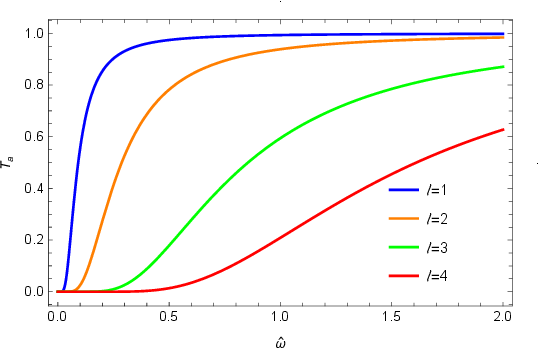}}\\
 \caption{For massless scalar perturbations, rigorous bounds on greybody factors are established 
 for different values of the multipole moment $l$ and metric parameter $\mho$, with parameter values $M=1$, $\bar{\bar{\omega}} = 1/3$~(radiation field), and~$P = -2$.}
    \label{fig:multi_graphs}\label{FIG.15}
     
  {\includegraphics[width=0.40\textwidth]{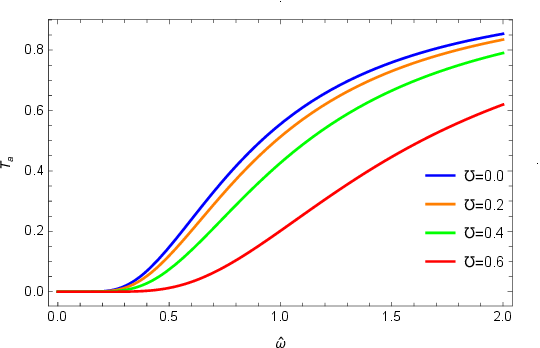}}\hfill
 {\includegraphics[width=0.40\textwidth]{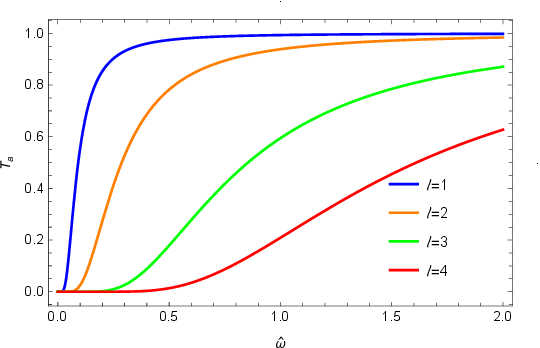}}\\
 \caption{With parameter values $M=1$, $\bar{\bar{\omega}} = 0$~(dust field), and~$P = - 2$, rigorous constraints on 
 greybody factors are obtained for various values of the multipole moment $l$ and metric parameter $\mho$ for massless scalar perturbations.}
    \label{fig:multi_graphs}\label{FIG.16}
 \end{figure}
\newpage
\begin{figure}

{\includegraphics[width=0.40\textwidth]{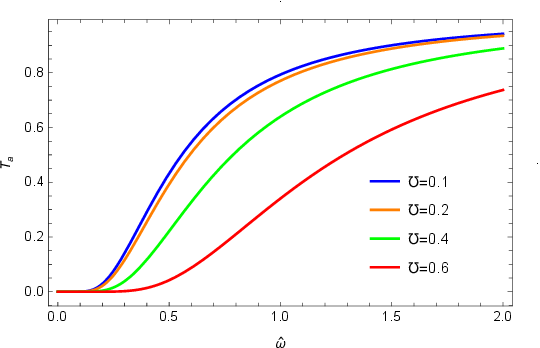}}\hfill
  {\includegraphics[width=0.40\textwidth]{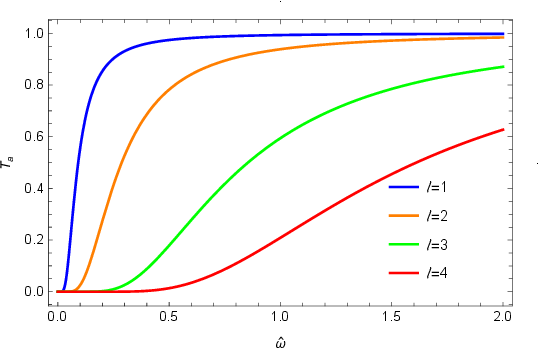}}\\
 \caption{Using parameter values $M = 1$, $\bar{\bar{\omega}} = - 2/3$~(quintessence field), and~$P = - 2$, rigorous constraints 
 on greybody factors are determined for massless scalar perturbations for varying values of the multipole moment $l$ and metric parameter $\mho$.}
    \label{fig:multi_graphs}\label{FIG.17}
    
{\includegraphics[width=0.40\textwidth]{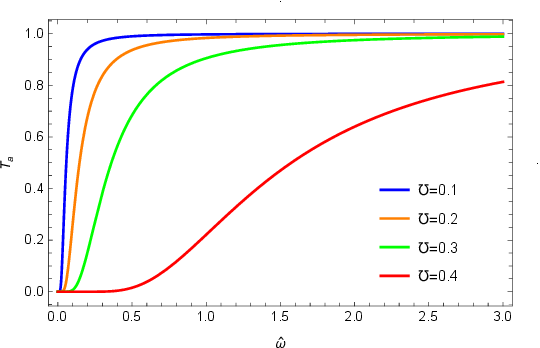}}\hfill
  {\includegraphics[width=0.40\textwidth]{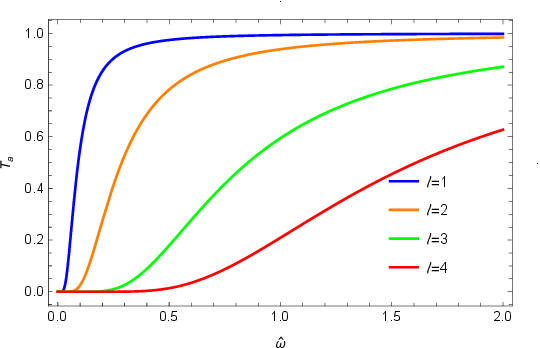}}\\
 \caption{Strict constraints on Greybody Components for massless scalar disturbances for a range of multipole 
 moment $l$ and metric parameter $\mho$ values, in which $M = 1$, $\bar{\bar{\omega}} = - 4/3$~(phantom field), and~$P = - 2$ are parameter values.}
    \label{fig:multi_graphs}\label{FIG.18}
    
{\includegraphics[width=0.40\textwidth]{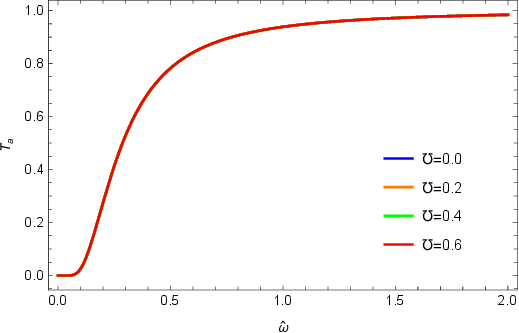}}\hfill
{\includegraphics[width=0.40\textwidth]{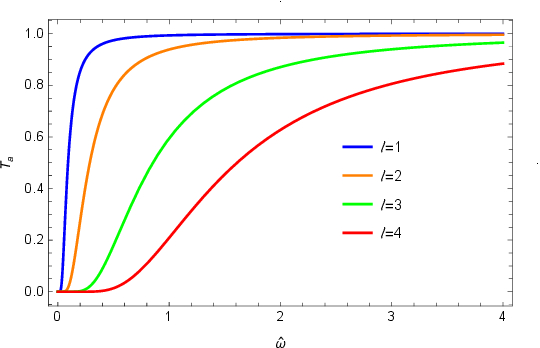}}\\
 \caption{Rigorous constraints on greybody factors are determined for massless scalar perturbations with 
 parameter values $M=1$, $\bar{\bar{\omega}} = - 1$~(cosmological constant field), and~$P = - 2$ for varying values of the multipole moment $l$ and metric parameter $\mho$.}
    \label{fig:multi_graphs}\label{FIG.19}

\end{figure} 
The bounds on GFs decrease as the values of the metric parameter $\mho$ increase. Figure $\textbf{19}$ 
indicates that the greybody boundaries are unaffected by changes in $\mho$.
According to the graphs, greybody bounds drastically drop as parameter $l$ rises, 
and parameter $l$ affects the greybody bounds more than the greybody factors discussed 
in the preceding subsection. This result indicates that BHs possess stronger barrier 
qualities in a gravity modified by quantum fluctuations and exhibit smaller greybody 
limitations compared to Schwarzschild BHs. Overall, the parameters of the BHs have a similar 
impact on the greybody bounds as discussed in the previous subsection regarding greybody factors. 

\section{Conclusion}

In this paper, we have explored the QNMs and GFs of BH solutions in gravity-modified 
by quantum fluctuations, with an emphasis on scalar and electromagnetic perturbations. According 
to our findings, the metric parameter $\mho$, the multipole moment $l$, and the state 
equation parameter $\bar{\bar{\omega}}$ greatly influence the QNMs and GFs. The parameter $\mho$ and the 
multipole moments $l$ are crucial when talking about the scattering and absorption characteristics of these black holes. \\
\textbf{Some key conclusions derived from our findings are:}
\begin{itemize}
\item Higher multipole moments \( l \) result in faster damping rates, increased QNM frequencies, 
and stronger potential barriers, enhancing BHs robustness against perturbations.
\item Additionally, the parameter value $\mho$ plays a crucial role and has a complex influence 
on the greybody components and the QNM spectrum. The damping rates and oscillation frequencies 
vary significantly at low values, but these effects become stable as $\mho$ increases.
\end{itemize}
The study reveals a significant difference between the GFs for BH with QFMG and those of a Schwarzschild BH.
We have discovered a correlation between smaller values of these parameters and higher GFs, which 
signify stronger interactions between the BH and incoming matter or waves.
Our investigation reveals that small positive values of \(\mho\) decrease the true frequencies 
of QNMs, while small negative values increase them. As \(\mho\) becomes extremely large, whether 
positive or negative, the QNMs approach constant values similar to those observed in a Schwarzschild BH.
Scalar perturbations consistently exhibit higher frequencies and shorter durations compared to 
electromagnetic perturbations, based on our analysis of scalar and vector perturbations. This 
suggests that the BH interacts more strongly with the scalar field. Consequently, the 
parameters \( l \) and \( \mho \) have measurable effects on the absorption and 
scattering behavior of the BH in relation to the greybody components.
We observe that the GFs increase as the parameter $\mho$ decreases, indicating that BHs with a 
lower $\mho$ are more effective at absorbing and scattering incoming radiation. Additionally, we 
calculated rigorous constraints on the GFs for scalar perturbations. The parameters influencing 
these rigorous bounds are the same as those that affect the GFs in previous cases.
It is important to note that ground-based gravitational wave detectors, like LIGO, are not capable 
of accurately detecting QNMs \cite{55}. In contrast, space-based gravitational wave detectors, such 
as LISA, are sensitive enough to detect QNMs from numerous significant sources, as demonstrated 
in \cite{55,56}. Therefore, we believe that LISA will soon be able to identify QNMs with greater accuracy.
These results offer valuable insights into how QFMG parameters affect the dynamics and observational 
characteristics of BHs. With the advancement of sensitive detectors like LISA and ongoing observations 
from the EHT, we may soon use these findings to further constrain QFMG and enhance our understanding 
of BH physics. Future experiments involving gravitational waves and BH shadows may reveal 
intriguing observational targets through the unique signatures of these parameters, particularly in QNM spectra and GFs.
Our recent work, along with the observational data on QNMs, enhances our understanding of gravity 
modified by quantum fluctuations. By combining the observational constraints on QNMs derived 
from LISA data with shadow constraints from the EHT, we may soon be able to evaluate the consistency and viability of this theory.

\end{document}